\documentclass[12pt, preprint]{aastex}
\usepackage{epsfig}
\usepackage{natbib}
\bibpunct{(}{)}{;}{a}{}{,} 
\shorttitle{gravitationally lensed QPO}
\shortauthors{Bursa et al.}

\def\d{{\rm d}}
\def\kB{k_{\rm B}}
\def\amu{m_{\rm u}}

\begin{document}
\title {The upper kHz QPO: a gravitationally lensed vertical oscillation}

\author{M. Bursa\altaffilmark{1,5},
M. A. Abramowicz\altaffilmark{2,5},
V. Karas\altaffilmark{3,5},
W. Klu\'zniak\altaffilmark{4,5}}
\altaffiltext1{Faculty of Mathematics and Physics, Charles University, 
CZ--180\,00 Prague, Czechia, bursa@sirrah.troja.mff.cuni.cz}
\altaffiltext2{Theoretical Physics, Chalmers University            
        S-412-96 G\"oteborg, Sweden, marek@fy.chalmers.se}
\altaffiltext3{Astronomical Institute, Academy of Sciences, 
CZ--141\,31 Prague, Czechia, vladimir.karas@cuni.cz}
\altaffiltext4{Institute of Astronomy, Zielona G\'ora University, 
        Wie\.za Braniborska,
        Lubuska 2, PL-65-265 Zielona G\'ora, and
        Copernicus Astronomical Centre, Bartycka 18, 
        PL-00-716 Warszawa, Poland, wlodek@camk.edu.pl}
\altaffiltext5{Visiting scientists at the UKAFF supercomputer facility, 
University of Leicester, England}

\begin{abstract}
We show that a luminous torus  in the Schwarzschild metric oscillating
along its  own axis gives rise to a periodically varying flux of
radiation, even though the  source of radiation is steady and
perfectly axisymmetric. This implies that the  simplest oscillation
mode in an accretion flow, axisymmetric up-and-down motion  at the
meridional epicyclic frequency, may be directly observable when it
occurs  in the inner parts of accretion flow around neutron stars and
black holes. The   high-frequency modulations of the X-ray flux
observed in low-mass X-ray binaries  at two frequencies (twin kHz
QPOs) could then be a signature of strong gravity  both because radial
and meridional oscillations have different frequencies in
non-Newtonian gravity, and because strong gravitational deflection of
light rays  causes the flux of radiation to be modulated at the higher
frequency.
\end{abstract}


\maketitle

\section{Highest frequency in accreting black holes and neutron stars}

The highest frequencies modulating the X-ray flux observed from accreting
neutron stars and black holes continue to attract attention because
their values are as high as those of orbital frequencies close
to the neutron star surface 
or to the circular photon orbit around a black hole.
The origin of the modulations, known as quasi-periodic 
oscillations (QPOs) because they are not quite coherent, 
still remains a major puzzle
(see van der Klis 2000 for a review).

Typically these QPOs come in pairs,
with the higher frequency (as high as 1.2 kHz for neutron stars
and 0.5 kHz for black holes) larger by about 50\%
than the lower frequency of the pair.
It has been recognized for some time that the HF QPOs may correspond to 
accretion disk oscillations not present in Newtonian $1/r$ gravity, 
e.g., modes trapped close to the maximum of the epicyclic 
frequency\footnote{The radial epicyclic frequency may also have a maximum for 
rapidly rotating Newtonian stars, and it is even possible that an innermost 
(marginally) stable circular orbit may then exist outside the stellar surface 
(Amsterdamski et al. 2002, Zdunik \& Gourgoulhon 2001).} 
(Wagoner 1999, Kato 2001). 
Non-axisymmetric modes have been preferred, as it was thought that a 
considerable degree of non-axisymmetry is a necessary condition for modulating 
the X-rays.
It has also been suggested that the HF QPO phenomenon
is caused by a non-linear resonance between
two modes of oscillation of the accretion disk or torus
(Klu\'zniak \& Abramowicz 2001, Abramowicz \& Klu\'zniak 2001).
In a resonance, there should be a rational ratio of frequencies, and indeed
the observed pairs of high-frequency QPOs in black holes
are in a 3:2 ratio (McClintock and Remillard 2003).
More recently, it has been specifically suggested that the higher frequency
corresponds to vertical oscillations of the accretion disk/torus
occuring at the meridional epicyclic frequency 
(Klu\'zniak \& Abramowicz 2002, 2003;
Klu\'zniak et al. 2004; Lee, Abramowicz \& Klu\'zniak 2004).
However, the mechanism of X-ray modulation remained a puzzle.

 Here, we show that 
gravitational lensing of the photon trajectories in Schwarzschild metric 
suffices to appreciably modulate the flux observed at infinity even if the 
source is symmetric about the axis of a black hole, provided  that it moves 
parallel to the symmetry axis. Specifically, we show that a toroidal source 
oscillating about the equatorial plane of the black hole, 
but otherwise steady, 
gives rise to a periodically modulated flux. If, in addition, the source is 
strongly variable at another frequency, the flux will strongly vary at two 
frequencies, with the power ratio depending on the inclination angle of the 
observer.

\section{Calculation of trajectories and observer flux}

In order to compute the amount of radiation coming from the source we
have  developed a~new three-dimensional ray-tracing code. Following
the method used by  Rauch~\& Blandford (1994) we integrate geodesic
and geodesic  deviation equations in the Schwarzschild
spacetime. Photon trajectories are  integrated backward in time from
the observer positioned at infinity at some  inclination angle~$i$
with respect to the~$z$ axis. At certain points along the  trajectory
the current position, momentum, time delay and magnification are
recorded. This information is then used to reconstruct each photon's
path and  calculate the total amount of incoming radiation.

The intensity observed at infinity is an~integration of 
the~emissivity $f$ over the~path length along geodesics and it can 
be written down as
\begin{equation}
I_{\rm obs}(t) = \int f(r,\,\theta,\,\phi,\,t-\Delta{t}) 
\sqrt{-g_{tt}}\, k^t \, g^4\, \d\lambda \;.
\end{equation}
The integration is along the~light ray  parametrized by an~affine 
parameter $\lambda$. Here, $k^t$ is the~time component of 
photon's \mbox{4-momentum}, $g$ is the~\mbox{red-shift} factor and 
$\Delta t$ is the~photon time delay.

\section{A luminous, vertically oscillating torus}
We consider an isolated, luminous,
optically thin and geometrically slender torus
around a non-rotating black hole of mass $M$. 
In this problem, all radii scale with $M$ 
(we use $G/c^2=1.5\,{\rm km}/M_\odot$), and all frequencies scale as
$1/M$.  For a convenient comparison with the observed frequencies
we chose $M=1.4M_\odot$.

The torus is assumed to be circular in cross-section and to oscillate
harmonically parallel to its axis.
In Sections 4 and 5,
we allow the same torus to execute radial oscillations
as well. The circle of its maximum pressure is at 
$\tilde{r}_0=10.8\,M(G/c^2)$ and
$\tilde{z}(t) = \delta\tilde{z}_0\,\sin(\omega_\theta t)$,
 in cylindrical co-ordinates related to the Schwarzschild co-ordinates
through
$\tilde{r} = r\,\sin\theta \;,
\tilde{z} = r\,\cos\theta$. Here, 
$\omega_\theta =\Omega_{\rm K}=\sqrt{GM/r^3}$
is the vertical epicyclic frequency, 
in Schwarzschild geometry equal to the Keplerian orbital frequency.
The amplitude of vertical motion is $\delta\tilde{z}_0=0.1M(G/c^2)$.
The intrinsic emissivity per unit co-moving volume of the torus is
held constant in time, as is its cross-sectional radius $R_0=1.5M(G/c^2)$.

In this {\sl Section}, the only time variation of the torus is in its
vertical position. 
We find that in spite of this, the flux observed at infinity clearly varies
at the oscillation frequency (Fig.~1). This is caused by
relativistic effects at the source (lensing and beaming),
and no other cause need be invoked to explain in principle
the highest-frequency modulation of X-rays in luminous black-hole 
binary sources.

\begin{figure*}[ht]
\label{axi}
\centering
\includegraphics[angle=-90, width=140mm]{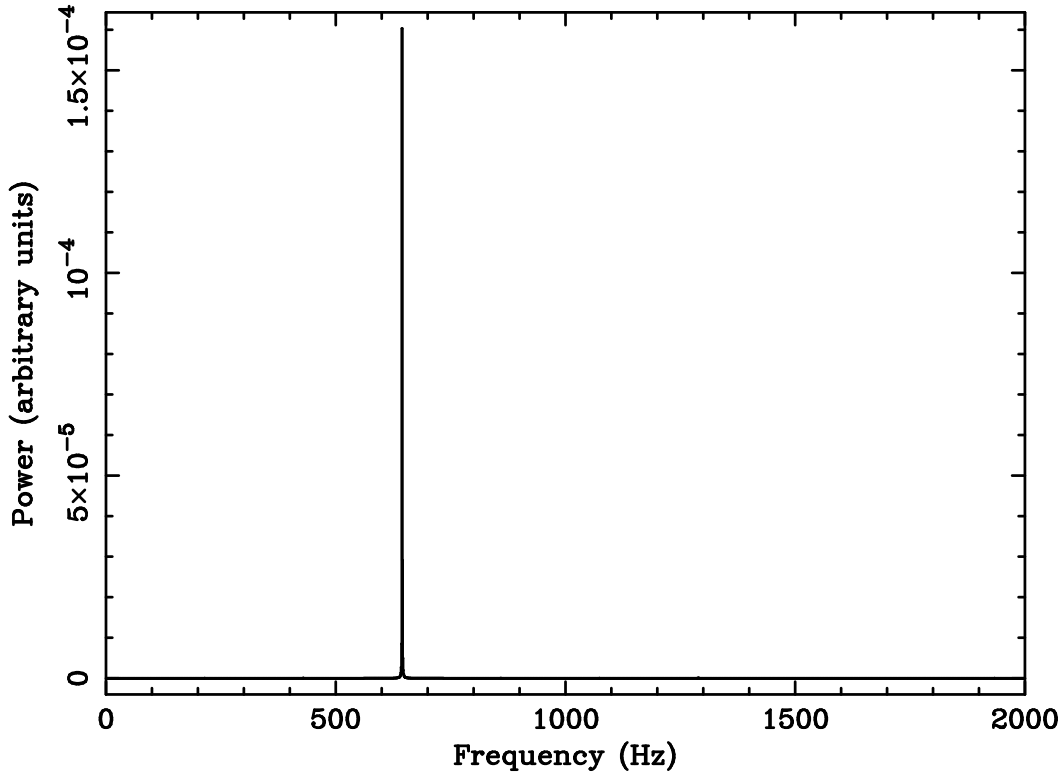}
\caption{\footnotesize The power spectrum of the radiation flux
of an axi-symmetric torus oscillating  vertically in Schwarzschild geometry.
The observer is at infinity, the inclination angle is $i=45^\circ$,
the power scale is arbitrary. To the level of parts per thousand,
all the harmonic content is in the frequency of vertical oscillation.
If, in addition, the torus is oscillating radially at a different
frequency, two strong Fourier components are seen (compare Fig.~3).
}
\end{figure*}

\section{Two oscillations of a torus}

Now consider radial oscillations of the torus under discussion.
Keeping the cross-sectional radius $R_0$ fixed we allow
the central circle of the torus to vary as
\begin{equation}
\tilde{r}(t) = \tilde{r}_0 + \delta\tilde{r}_0\,\sin(\omega_r t),
\end{equation}
simultaneously with 
\begin{equation}
\tilde{z}(t) = \delta\tilde{z}_0\,\sin(\omega_\theta t).
\end{equation}
This results in a periodic change of volume of the torus.
Because the optically thin torus is assumed to be filled with 
a polytropic gas radiating by 
bremsstrahlung cooling, there is a corresponding change of luminosity,
with a clear periodicity at $2\pi/\omega_r$.
With our choice of $\tilde{r}_0$, we have $\omega_r={2\over 3}\omega_\theta$.

The luminosity variations will depend on the properties of the torus.
We take the emissivity in the local frame to be
$f\propto\rho^2\,T^{\frac12}$,
with
$T = K\,\rho^{\gamma-1} {\mu\amu}/{\kB}$,
where
 $\gamma=\frac53$, $\mu=\frac74$, $\amu$ and $\kB$ are polytropic 
index, molecular weight, atomic mass unit and the~Boltzmann constant, 
respectively.
To approximate the properties of an oscillating torus, we actually
took the equipotential structure obtained by Taylor expanding 
in the $\tilde{z}$ direction equilibrium solutions 
of the relativistic Euler equation (Abramowicz et al. 1978) 
of a torus with uniform angular momentum
$\ell(\tilde{r}) = \ell_{\rm K}(\tilde{r}_0) = 
{\sqrt{M\,\tilde{r}_0^3}}/{(\tilde{r}_0-2M)}$, so that 
$\rho = \left[ \frac{\gamma-1}{K\gamma} 
\left({\rm e}^{\Delta W} - 1\right)\right]^{{1}/{\gamma-1}}$,
and $\Delta{W} = ({R_0^2-R^2})/[2\,\tilde{r}_0^2(\tilde{r}_0-3M)]$.
The net effect is best displayed as the power spectrum at infinity
of a flux of photons propagating in Minkowski space, i.e., one
obtained neglecting the relativistic effects responsible for
modulating the flux at the vertical epicyclic frequency. The resulting power
spectrum of a torus oscillating with amplitudes
$\delta\tilde{z}_0=\delta\tilde{r}_0=0.1M(G/c^2)$
about a Newtonian point mass $M$ is shown in Fig.~2.

 \begin{figure*}[ht]
\label{flat}
\centering
\includegraphics[angle=-90, width=140mm]{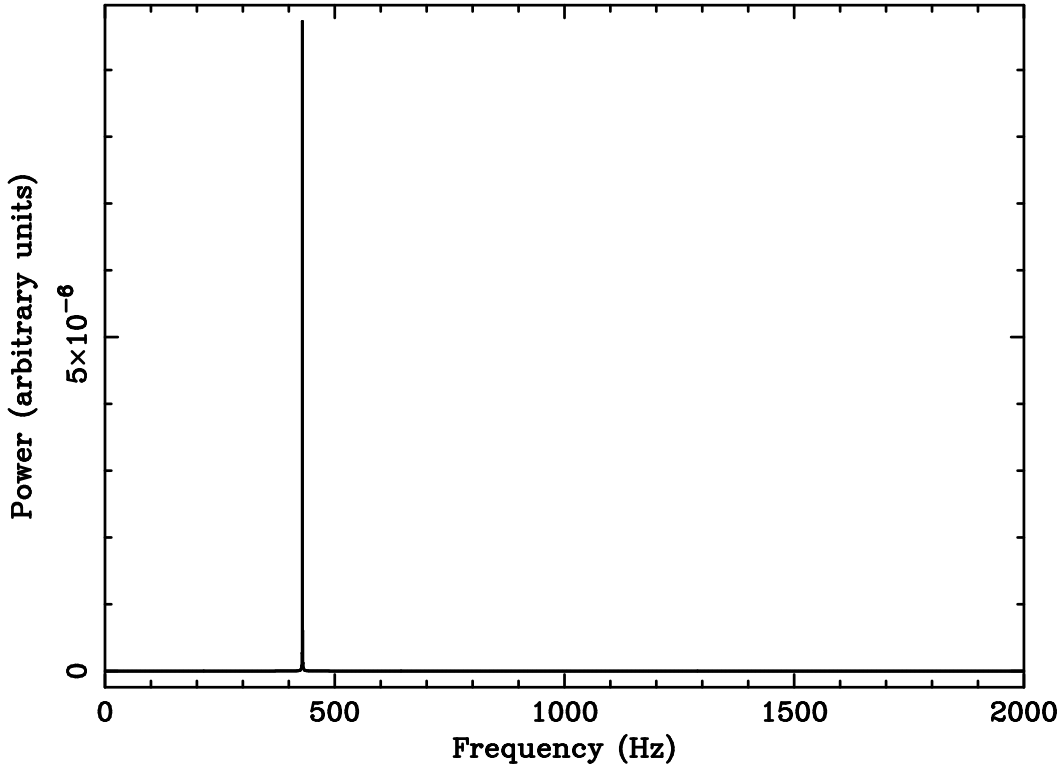}
\caption{\footnotesize The power spectrum of an axi-symmetric
 torus oscillating radially and vertically about a Newtonian point mass
(the power scale is arbitrary).
The calculation is performed exactly as for the other figures,
except that
 the light trajectories are computed in Minkowski space, and Doppler beaming
has been switched off. The radial oscillations result in a harmonic change
in volume, and hence also in the luminosity, of the torus.
Note that all the power is at the frequency of radial oscillation.
Without lensing (or Doppler beaming) there is no modulation
at the frequency of vertical motion (compare Figs.~1,~3,~4).
}
\end{figure*}

\section{Modulation of the light curve}
We have computed the light trajectories for several inclination
angles in Schwarzschild geometry
for the oscillating torus described in Section~4.
In all cases, two strong periodic components are clearly seen
in the light curves and in the power spectra, at the two oscillation
frequencies $\omega_r$ and $\omega_\theta$.
The relative power in the two components depends on the inclination
angle and the amplitude of oscillation (Figs. 3, 4).

\begin{figure*}[ht]
\label{big}
\centering
\includegraphics[angle=0, width=14cm]{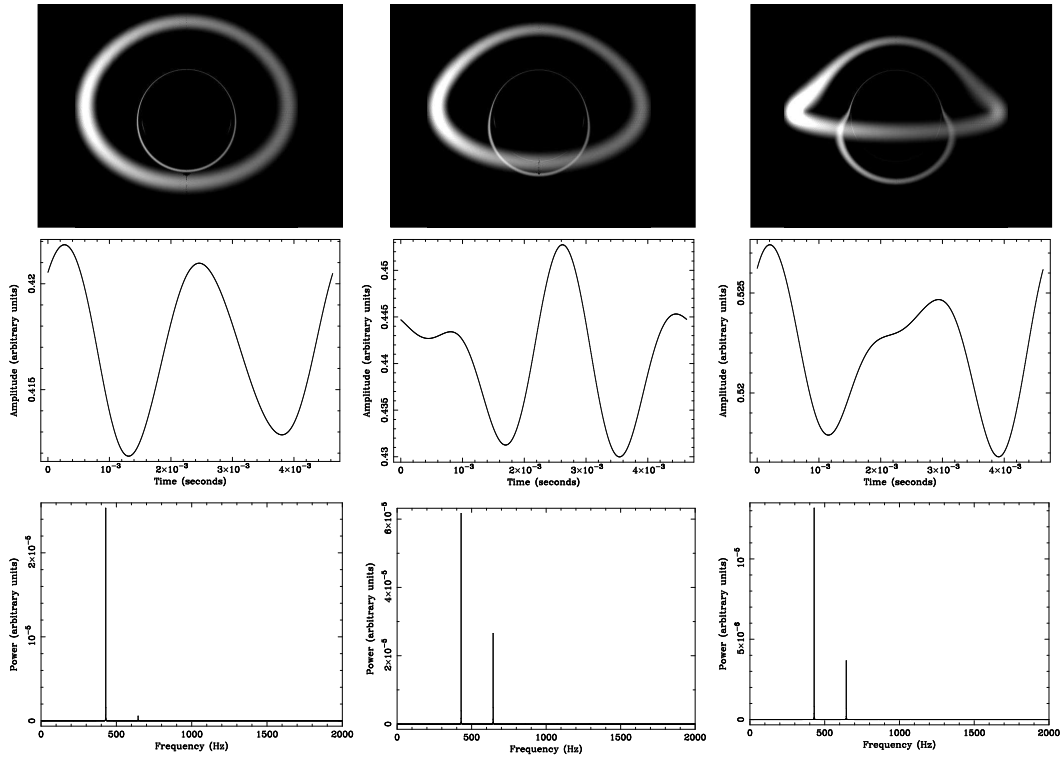}
\caption{Results of numerical simulations of the~oscillating torus
in Schwarzschild geometry. The 
equilibrium distace of the torus $\tilde{r}_0=10.8M$,
its cross-section radius is $R_0=1.5M$,
the oscillation amplitudes are
$\delta\tilde{z}_0=\delta\tilde{r}_0=0.1M$,
and frequencies $\omega_\theta=\Omega_{\rm K}$,
$\omega_r=\frac23\Omega_{\rm K}$. 
(Top):--snapshots of an~instant image, as
viewed by a~distant observer,
(middle):--the~computed light curve, and
(bottom):--the~corresponding power spectrum,
for three different viewing angles, $i=45^\circ$  (left), 
$60^\circ$ (middle) and $80^\circ$ (right).
}
\end{figure*}

\begin{figure*}[ht]
\label{small}
\centering
\includegraphics[angle=0, width=140mm]{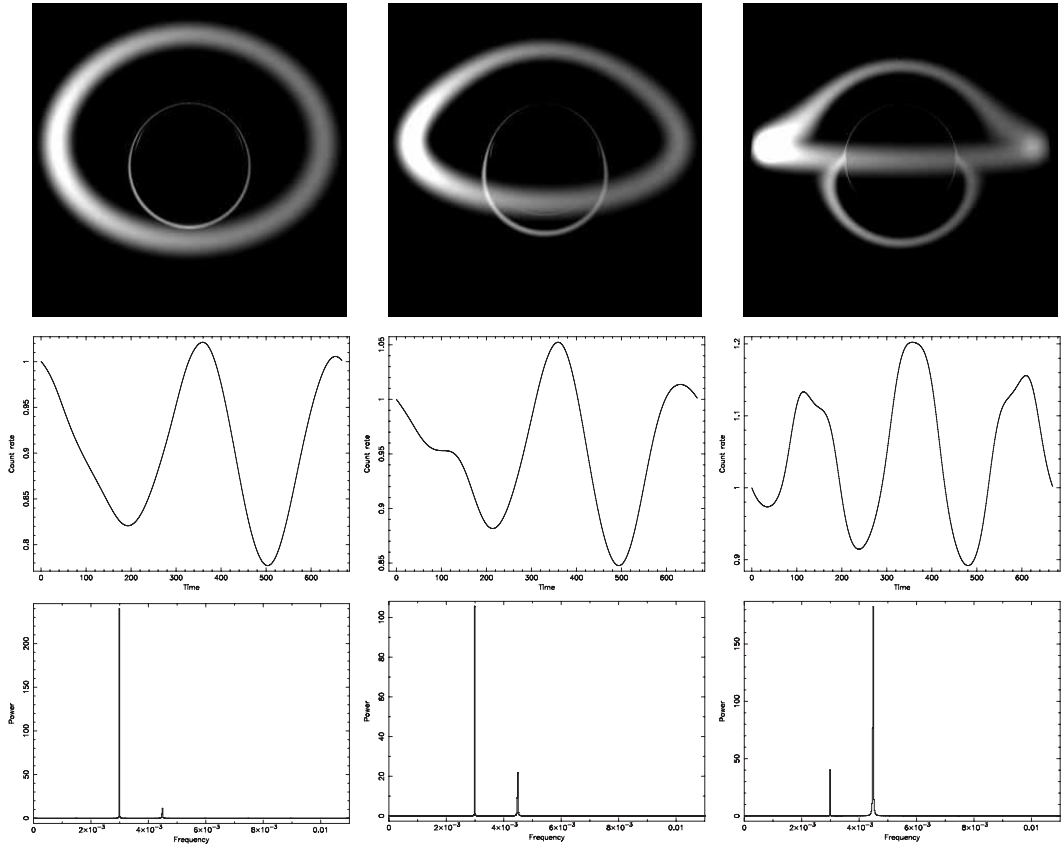}
\caption{Same as Fig.~3, but for 
$R_0=2M$,
$\delta\tilde{z}_0 =\delta\tilde{r}_0 = 1M$ 
and inclinations
 $i=45^\circ$, $65^\circ$, and $85^\circ$.
}
\end{figure*}

\section{Discussion}

We have shown that gravitational lensing at the source
will modulate the flux received at infinity
from an axially symmetric emitter oscillating about the equatorial plane
of a black hole.
We also find,
that in a torus executing simultaneous oscillations in the radial and vertical
directions at frequencies $\omega_r$,
and $\omega_\theta=(3/2)\omega_r$, as expected in the parametric resonance
model (Klu\'zniak and Abramowicz 2002), both frequencies will show up in the
power spectrum, with no other (e.g., harmonic) strong components.
The lower of the frequencies may reflect changes in the emissivity
of the torus, but the presence of the upper frequency is explained
by effects of relativity alone. Strong-field gravity may thus have
two signatures in the observed fast variability of black hole emission.
First, it is responsible for the presence of two frequencies
$\omega_\theta\ne\omega_r$, where in Newtonian gravity there is only one
($\omega_\theta=\omega_r=\Omega_{\rm K}$).
Second, it is responsible for modulation of the light curve (Figs.~3,4),
where in Newtonian gravity there was none at the
frequency of the vertical oscillations (Fig.~2).

\begin{acknowledgements}

This work was supported by the UK Astrophysical Fluids Facility
(UKAFF) supported through the ``European Community --- Access to
Research Infrastructure action of the Improving Human Potential
Program'', by the~European Commission through the grant number
\mbox{HPRI-CT-1999-00026} (the TRACS Programme at EPCC), by
the~GA\v{C}R Grant \mbox{No.\,205/03/0902} and by the Polish KBN grant 2P03D01424. 
\end{acknowledgements}


\begin{thebibliography}{}

\bibitem []{} Amsterdamski, P., Bulik, T., Gondek-Rosi{\'n}ska, D.,
               Klu{\'z}niak, W. 2002, A\&A, 381 L21

\bibitem []{} Abramowicz M.A., Jaroszy{\'n}ski M., Sikora M. 1978,
              A\&A 63, 221

\bibitem []{} Abramowicz M.A. \& Klu{\'z}niak W. 2001,
              A\&A 374, L19

\bibitem []{} Abramowicz M.A. \& Klu{\'z}niak W. 2003,
              in X-Ray Timing 2003: Rossi and Beyond, P. Kaaret,
              J. Swank, eds. in press, astro-ph/0312396

\bibitem []{} Kato, S. 2001, PASJ 53, 1

\bibitem []{} King, A. R., Davies, M. B., Ward, M. J., 
              Fabbiano, G.,  Elvis, M. 2001,                
              ApJ 552, L109
  
\bibitem []{} Klu{\'z}niak W., Abramowicz M.A. 2001, 
              Acta Physica Polonica B 32, 3605
              [http://th-www.if.uj.edu.pl/acta/vol32/t11.htm]
              
\bibitem []{} Klu{\'z}niak W., Abramowicz M.A. 2002,
              astro-ph/0203314
             
\bibitem []{} Klu{\'z}niak W., Abramowicz M.A. 2003,
              12th Workshop on General Relativity and Gravitation, 
              \\(Tokyo: Tokyo University Press), astro-ph/0304345 

\bibitem []{} Klu{\'z}niak, W., Abramowicz, M. A., Kato, S., Lee, W. H., 
             Stergioulas, N. 2004, ApJ 603, L89

\bibitem []{} Lee, W. H., Abramowicz, M. A., Kluzniak, W. 2004,
              ApJ, 603 L93

\bibitem []{} McClintock J.E, Remillard R.A. 2003,
              astro-ph/0306213 v.2

\bibitem []{} Rauch K. P., Blandford R. D., 1994, ApJ 421, 46

\bibitem []{} van der Klis M. 2000, 
              Ann. Rev. of A\&A 38, 717

\bibitem []{} Wagoner R.V.  1999, Phys. Rev. 311, 259

\bibitem []{}  Zdunik, L., Gourgoulhon, E., 2001, PhRvD, 63h7501

\end{thebibliography}
\end{document}